# Two-Dimensional Charge Localization at the Perovskite Oxide Interface

Chi Sin Tang,[1,2,3,†] Shengwei Zeng,[4,†] Caozheng Diao,[3] Jing Wu,[2] Shunfeng Chen,[1] Mark B. H. Breese,[3,4] Chuanbing Cai,[1] Ariando Ariando,[4,*] Andrew T. S. Wee,[4,5,*] Xinmao Yin,[1,*]

[1] Shanghai Key Laboratory of High Temperature Superconductors, Physics Department, Shanghai University, Shanghai 200444, China
[2] Institute of Materials Research and Engineering, A∗STAR (Agency for Science, Technology and Research), 2 Fusionopolis Way, Singapore, 138634 Singapore
[3] Singapore Synchrotron Light Source (SSLS), National University of Singapore, Singapore 117603
[4] Department of Physics, Faculty of Science, National University of Singapore, Singapore 117542
[5] Centre for Advanced 2D Materials and Graphene Research, National University of Singapore, Singapore 117546

†These authors contributed equally to this work.
*To whom correspondence should be addressed.
E-mail: yinxinmao@shu.edu.cn, phyarian@nus.edu.sg, phyweets@nus.edu.sg

**Abstract**

The effects of atomic-scale disorder and charge (de)localization holds significant importance, and they provide essential insights in unravelling the role that strong and weak correlations play in condensed matter systems. For perovskite oxide heterostructures, while disorders introduced via various external stimuli have strong influences on the (de)localization of the interfacial two-dimensional (2D) electrons, these factors alone could not fully account for the system's charge dynamics where interfacial hybridization holds very strong influence. Here, we determine that the displaced 2D free electrons are localized in the specific hybridized states at the LaAlO$_3$/SrTiO$_3$ (LAO/STO) interface. This experimental study combines both transport measurements and temperature-dependent X-ray absorption spectroscopy and suggests the localization of 2D electrons can be induced via temperature reduction or ionic liquid gating and it applies to both amorphous and crystalline interfacial systems. Specifically, we demonstrate that interfacial hybridization plays a pivotal role in regulating the 2D electron localization effects. Our study resolves the location where the 2D electrons are localized and highlights the importance of interfacial hybridization and opens further scientific investigation of its influence on 2D charge localization in oxide heterointerfaces.

**Main Text**

The customization of nanoscale disorder effects in condensed matter systems can come in the form of physical (e.g., defects, distortions) and chemical disorders (e.g., impurities and dopants). Disorder control serves to produce localized states leading to charge localization via symmetry breaking processes[1,2]. The effects of disorder and charge (de)localization in correlated systems are important in Condensed Matter Physics. They provide insights to unravel the role of strong and weak correlations between electrons in systems including unconventional cuprate superconductors[3-6]. When (de)localization is translated to low-dimension systems, unique phenomena would arise due to quantum confinement. Besides, localized electronic states and charge separation play an equally crucial role in applications relating to semiconductors and photovoltaics[7-9].

In two-dimensional electron gas (2DEG) at perovskite oxide interfaces including the quintessential $LaAlO_3$/$SrTiO_3$ (LAO/STO)[10], studies have reported that the introduction of external stimuli including electric field effect[11,12], LAO stoichiometric control[13,14], oxygen partial pressure, temperature regulation[15,16] and laser irradiation[17,18] have strong influences on the (de)localization of the interfacial 2D-electrons. However, these external stimuli to introduce physical/chemical disorders in the respective LAO/STO layers alone could not account for the interfacial charge (de)localization effects. Hence, other essential contributing factors to the disorder and charge (de)localization must be considered. Besides, it is crucial to determine where the 2DEG are localized when the LAO/STO interfacial charge concentration is reduced. Meanwhile, although interfacial orbital hybridization plays an important role in dictating the distinct behaviour of the 2DEG, it has received insufficient attention for further investigation in this complex interfacial system.

Temperature-dependent X-ray Absorption Spectroscopy (XAS) is an element-specific characterization technique that offers opportunities to investigate the evolution in the electronic structures of quantum systems, interfacial effects, and any arising orbital hybridization[19-22]. Its surface sensitivity makes it an ideal experimental tool to characterize ultra-thin film and interfacial systems. It is therefore possible to locate the displaced 2DEG at the LAO/STO interface via XAS and to unravel the role that interfacial hybridization plays in the 2D charge localization phenomenon.

In this letter, we report a significant drop in 2D electron concentration at the LAO/STO interface induced via temperature reduction and ionic liquid (IL) gating. It is then determined that this significant reduction in 2DEG is due to the localization of charges at the interfacial $O2p$-$Ti3d(e_g)$ and $O2p$-$Sr4d$ states ($O2p$ orbitals belonging to the LAO layer while the $Ti3d(e_g)$ and $Sr4d$ orbitals to the STO layer). We demonstrate that interfacial hybridization between the LAO and STO layers are pivotal in regulating the 2D electron localization processes (Fig. 1(a)). This comprehensive study combines transport measurements and temperature-dependent XAS over a wide temperature range on LAO/STO systems before and after IL-gating with different LAO film thicknesses. Our study provides important clues on how interfacial electrodynamics can lead to inter-layer hybridization and charge localization in both amorphous and crystalline oxide heterostructures. Furthermore, it provides new opportunities to unveil the mechanisms leading to 2D charge localization in complex heterostructures and could potentially be exploited in modulating 2D charge carrier properties as a means of device control in oxide-based interfaces.

The LAO/STO interfaces were derived by depositing both amorphous and crystalline LAO on [100]-oriented STO substrates using pulsed laser deposition (details in Supplementary). For the samples synthesized for undergoing transport and XAS characterization are shown in Figs. 1(b) and (c), respectively. Temperature-dependent XAS measurements were performed at the Soft X-ray–ultraviolet (SUV) beamline in the Singapore Synchrotron Light Source (see Supplementary).

Figures 2(a)–(c) compare the sheet resistance, $R_s$, carrier density, $n_s$, and Hall mobility, $\mu_H$, with respect to temperature for the amorphous 4.0nm-LAO/STO device characterized before and after IL-gating (See Supplementary). The pristine 4.0nm-LAO/STO exhibited metallic features with decreasing sheet resistance alongside a corresponding dip in carrier concentration as temperature decreases (Figs. 2(a) and (b), respectively). The temperature-dependent transport properties of oxygen-deficient crystalline 4.0nm-LAO/STO display similar metallic behaviour where the interfacial electrons are also partially missing at low temperature (Fig. S3).

After IL-gating, the amorphous 4.0nm-LAO/STO continue to display metallic behavior with similar temperature-dependent sheet resistance, $R_s$-$T$, behavior with a slightly lower $R_s$ at low temperature but a higher $R_s$ at high temperature. However, its carrier density, $n_s$, is significantly reduced and becomes virtually temperature-independent (Fig. 2(b)). Compared to the sample in its pristine state, the amorphous LAO/STO displays higher $\mu_H$ at low temperature (Fig. 2(c)). These data suggest that the interfacial 2D electrons are partially displaced after IL-gating takes place in the amorphous LAO/STO.

To locate the displaced delocalized electrons in both the amorphous and crystalline 4.0nm-LAO/STO samples in their pristine state at low temperature, it is crucial to study how the system's interfacial electronic structure evolves with temperature. Temperature-dependent XAS characterization is conducted to alongside other interfaces to compare their electronic structures. Particularly, analysis of the O K-edge spectra characterizes the interfacial hybridization between the O2$p$ and the metallic orbital states–thereby providing important information of the system's electronic structure in terms of the unoccupied metallic density of states.

Figure 2(d) shows the temperature-dependent O K-edge XAS spectra of the amorphous 4.0nm-LAO/STO interface in its pristine state where the spectra are reproducible (see Supplementary). The general shape of the spectra is consistent with previous studies focusing on 2D electronic structure at the LAO/STO interface where the main spectral features are labelled peaks A, B and C[23,24]. Peak A (~531.9 eV) is attributed to the O2$p$—Ti3$d$($t_{2g}$) hybridization [25-27]. Its intensity decreases with temperature dropping to 100 K and remains stable thereafter. This is consistent with the electron density reduction near the Fermi level (O2$p$—Ti3$d$($t_{2g}$) states) indicated by the carrier density measurement (Fig. 2(b)). Furthermore, this abrupt drop in electron density at 100 K coincides with the tetragonal-to-cubic structural phase transition of the STO substrate at ~105 K (to be discussed thereafter). The decreasing intensity of peak A corresponding to the weakening O2$p$—Ti3$d$ ($t_{2g}$) hybridization is similar to that of etched STO (Fig. 3(c)). The consistency of the temperature changes in the O2$p$—Ti3$d$($t_{2g}$) hybridized states with that of etched STO (presented thereafter) and the oxygen vacancies show that they are not the main factors for the reduced interfacial electrons with decreasing temperature.

We next consider the features in the 532—539 eV region corresponding to the overlapping La5$d$ and O2$p$ orbital hybridization, the O2$p$—Ti3$d$($e_g$) and O2$p$—Sr4$d$ hybridizations at the 2D interface where the O2$p$ orbitals belong to the LAO layer while the Ti3$d$($e_g$) and Sr4$d$ orbitals are contributions from the STO substrate layer[23,24,28]. Features B and C located within this spectral region display similar temperature-dependent behavior as peak A. Particularly, while feature B (~535.5 eV) registers the highest intensity at 300 K, its intensity decreases with falling temperature and eventually forms an overlapping shoulder with marginal changes between 35 and 100 K. Apart from an intensity drop and feature sharpening with decreasing temperature, peak C also registers a blue shift from ~537.6 eV at 300 K to ~538.3 eV at 35 K. To compare the intensity changes of the respective peaks in pristine 4.0nm-LAO/STO, a temperature-dependent differential, $\Delta I$ (where $\Delta I=I(E,T)-I(E,35K)$), analysis is performed as shown in Fig. 2(e). It indicates that while temperature variations are large above 100 K, these changes become significantly subdued between 35 and 100 K.

To contrast the observation in the pristine 4.0nm-LAO/STO heterostructure where the 2DEG population is inherently large, a similar XAS characterization is conducted for the LAO/STO sample after IL-gating has been performed on it. Fig. 2(f) displays the temperature-dependent O K-edge spectra of the 4.0nm-LAO/STO after IL-gating. While the O K-edge spectra are generally like its former pristine state, its temperature-dependent trend is considerably and irreversibly subdued (see Supplementary). This is more distinctly observed via the intensity differential, $\Delta I$ (Fig. 2(g)). While the temperature-dependent behavior of the gated 4.0nm-LAO/STO is significantly subdued than its pristine counterpart, the similar temperature-dependent trends particularly in cases of peaks A and B highlights the influence that the structural transition of the STO substrate has on the interfacial 2DEG.

As crystalline 4.0nm-LAO/STO registers similar transport property as its amorphous counterpart in its pristine state (see Figs. 2(a)-(c) and S3), O K-edge XAS characterization is conducted to investigate if the crystalline interfacial electronic structure evolves the same way with temperature. The O K-edge spectra for crystalline 4.0nm-LAO/STO in its pristine state before gating are displayed in Fig. 3(a). While the spectral shapes are relatively different due to its crystalline structure, the main features attributed to the interfacial orbital hybridization observed in its amorphous counterpart are present at similar energy positions and are labelled A", B" and C" where they denote the O2$p$—Ti3$d$($t_{2g}$), O2$p$—Ti3$d$($e_g$) and O2$p$—Sr4$d$ 2D interfacial hybridizations, respectively. Temperature differential analysis, $\Delta I$ (compare Figs. 2(e) and 4(b)) show that these features have similar temperature-dependent behaviour with weakened intensity with decreasing temperature. These are clear indications that temperature effects are applicable to both the amorphous and crystalline interfaces.

To confirm the 2D electronic behavior at the LAO/STO interface and to effectively distinguish between the interfacial electronic structures from lattice effects of the bulk layer and any arising effects induced by oxygen vacancies, a temperature-dependent investigation is also conducted on bare STO substrate. Fig. 3(c) displays the temperature-dependent O K-edge spectra of etched (Ti-O terminated) STO. It is an effective comparison with pristine 4.0nm-LAO/STO due to its high oxygen vacancy concentration[29]. Nevertheless, note that the spectra feature between these 2 systems are largely different. This provide substantial proof that the XAS spectra of both amorphous and crystalline 4.0nm-LAO/STO interfaces as further substantiated by the characterization of 50nm-LAO/STO presented thereafter.

For etched STO, four main features: A*, B*, C* and D*, are clearly observed in the O K-edge spectra with varying intensities with temperature. Peak A* (~531.9 eV) is attributed to the O2$p$—Ti3$d$($t_{2g}$) hybridization. while Peak B* (~534.4 eV) is ascribed to the O2$p$—Ti3$d$($e_g$) hybridization. Peaks C* (~539 eV) and D* (~544.3 eV) are attributed to the O2$p$—Sr4$d$ and O2$p$—Ti4$sp$ hybridization, respectively. These transition peaks are consistent with other studies[25-27].

The intensity drop in peaks A* and B* below 100K is attributed to the second-order tetragonal-to-cubic structural phase transition that STO undergoes at ~105 K[30-32]. This structural change influences the Ti-O hybridization[33,34] as seen in the intensity decrease of peaks A* and B*, indicating a diminishing Ti3$d$-O2$p$ hybridization strength (See Supplementary)[35]. Meanwhile, peaks C* and D* show an opposite temperature trend where their intensities are significantly maximized at 35K (See Supplementary).

Comparing the temperature trends between pristine 4.0nm-LAO/STO heterointerfaces (Figs. 2(d,e) and 3(a,b), respectively) and etched STO (Fig. 3(c)), the temperature behavior especially in the 532—539 eV range corresponding to features B* and C* (B and C in pristine 4.0nm-LAO/STO) are entirely contrary. As mentioned, both the pristine 4.0nm-LAO/STO and etched STO sample have high concentration of oxygen vacancies[29]. The contrary temperature trends of these two systems are clear indications that the temperature-dependent behavior of pristine 4.0nm-LAO/STO is not mediated by oxygen vacancies as registered in etched STO. Instead, these changes must be ascribed to the electronic structures and the reduction in 2D interfacial electron density. The anomalous intensity dip in peaks B and C with decreasing temperature in pristine 4.0nm-LAO/STO (Fig. 2(d)) indicates the filling of unoccupied interfacial 2D electronic states by electrons. This strongly suggests that the displaced interfacial electrons indicated by transport measurements (Fig. 2 and S3) have been localized in the O2$p$—Ti3$d$($e_g$) and O2$p$—Sr4$d$ states at low temperature. Therefore, the contribution to peaks B and C of both amorphous and crystalline 4.0nm-LAO/STO interfaces can be attributed to three possible components: 1. The hybridized states solely from the STO layer; 2. The hybridized states solely from the LAO layer (to address thereafter in Figs. 3(e) and (f)); 3. The interfacial hybridization between the O2$p$-orbitals of the LAO layer and the Ti3$d$($e_g$) and Sr4$d$-orbitals of the STO layer. Based on the O K-edge temperature trends of 4.0nm-LAO/STO interfaces presented in Figs. 2(d)-(g) and 3(a)-(b), and that of the 50nm-LAO/STO presented thereafter in Figs. 3(e) and (f), we conclude that temperature-regulated interfacial hybridization is the primary reason for the low-temperature charge localization in pristine 4.0nm-LAO/STO interfaces. The significant drop in 2D mobile interfacial electrons at low temperature registered in the transport measurements is due to their localization in the O2$p$-Ti3$d$($e_g$) and O2$p$-Sr4$d$ interfacial hybridized states where the O2$p$-orbitals (LAO layer) and the Ti3$d$($e_g$) and Sr4$d$ orbitals (STO substrate). This phenomenon applies to both amorphous and single-crystalline LAO/STO systems[36].

Further characterization is conducted on a pristine amorphous 50nm-LAO/STO to confirm the interfacial features of the 4.0nm-LAO/STO interfaces. With the sheer thickness of the 50nm-LAO layer, the temperature-dependent O K-edge spectra (Fig. 3(e)) and its temperature differential, $\Delta I$, (Fig. 3(f)) is entirely from the LAO layer[23,27]. Hence, the absorption features just above ~530 eV that corresponding to the O2$p$-Ti3$d$ hybridization are no longer visible. Instead, it is replaced by the O2$p$-Al3$p$ ($X$: ~531 eV), O2$p$-La5$d$ $t_{2g}$ ($Y$: ~534.8 eV) and $e_g$ ($Z$: ~537.4 eV) hybridization features, respectively[23,24]. Unlike the temperature-dependent behavior of the pristine amorphous and crystalline 4.0nm-LAO/STO (Figs. 2(d) and 3(a), respectively), the temperature-dependence of 50nm-LAO/STO is considerably subdued (Fig. 3(f); see Supplementary). This again substantiates our claim that the XAS features of the 4.0nm-LAO/STO systems are attributed to the interfacial effects.

To account for the underlying mechanism dictating the interfacial effects in the pristine 4.0nm-LAO/STO systems (Figs. 2(d,e) and 3(a,b), respectively), higher oxygen vacancy concentration in the STO layer leads to higher 2D charge population at ambient temperature. This strengthens the electron screening effect at the STO surface[37], which invariably weakens the electronic correlations at the LAO/STO interface. However, as a previous study demonstrated that the electron screening effect will weaken with decreasing temperature[37], this strengthens the system's electronic correlation and disorder[22,37]. With pre-existing interfacial hybridization, the strengthened electronic correlation and disorder bring about the localization of 2DEG in the interfacial O2$p$—Ti3$d$($e_g$) and O2$p$—Sr4$d$ hybridized states, thereby reducing the interfacial charge population. This charge localization effect can be reversed through thermal excitation as seen in both the transport and XAS measurements (Figs. 2(b), (d), (e) and 3(a), (b); See Supplementary).

Conversely, after IL-gating significantly and irreversibly reduces the interfacial 2D charge population in the amorphous 4.0nm-LAO/STO and can no longer be regulated by temperature (Figs. 2(b), (f) and (g)). The O K-edge spectra of pristine and IL-gated 4.0nm-LAO/STO at 300K (Fig. 3(d)) shows a marked intensity decrease in O2$p$—Ti3$d$($e_g$) and O2$p$—Sr4$d$ hybridized states after IL-gating. Thus, indicating that the 2D electrons have been irreversibly localized in these interfacial hybridized states after IL-gating. The reduction in free carriers by IL-gating significantly weakens the interfacial screening effect even at 300 K. Hence, the electronic correlations prevail throughout the entire temperature range, keeping the interfacial charges irreversibly localized in the hybridized states as indicated by the transport and XAS data (Figs. 2(b),

(f) and (g)). While this phenomenon is overlooked in previous studies involving bulk STO [38-40], our results clearly demonstrate the importance of electronic correlations extending beyond the STO layer into the LAO/STO interface.

In conclusion, we have demonstrated that the significant drop in 2DEG concentration at the LAO/STO interface brought about via temperature reduction or IL-gating is attributed to the electron localization at the interfacial hybridized states. These experimental results of the 2D electron localization imply that interfacial hybridization is an important contributing factor and reopens the investigation on how it influences the 2D charge localization in oxide heterointerfaces. While such temperature-dependent charge localization properties are observed in both crystalline and amorphous LAO/STO interfaces – indicating that these robust interfacial electronic properties are present generally in both systems, it is of considerably greater ease to synthesize amorphous samples even at room temperature which makes it compatible with established semiconductor fabrication processes[41-43]. Therefore, it is advantageous to consider this study not only at the fundamental research level, but this study also opens a new avenue for constructing and controlling high-mobility oxide interfaces in practical device applications.

**Acknowledgment:**
Chi Sin Tang and Shengwei Zeng contributed equally to this work. This research is supported by the Agency for Science, Technology, and Research (A*STAR) under its Advanced Manufacturing and Engineering (AME) Individual Research Grant (IRG) (A1983c0034) and the National Research Foundation, Singapore, under the Competitive Research Programs (CRP Grant No. NRF-CRP15-2015-01). The authors would like to acknowledge the Singapore Synchrotron Light Source for providing the facility necessary for conducting the research. The Laboratory is a National Research Infrastructure under the National Research Foundation, Singapore. Any opinions, findings and conclusions or recommendations expressed in this material are those of the author(s) and do not reflect the views of National Research Foundation, Singapore.

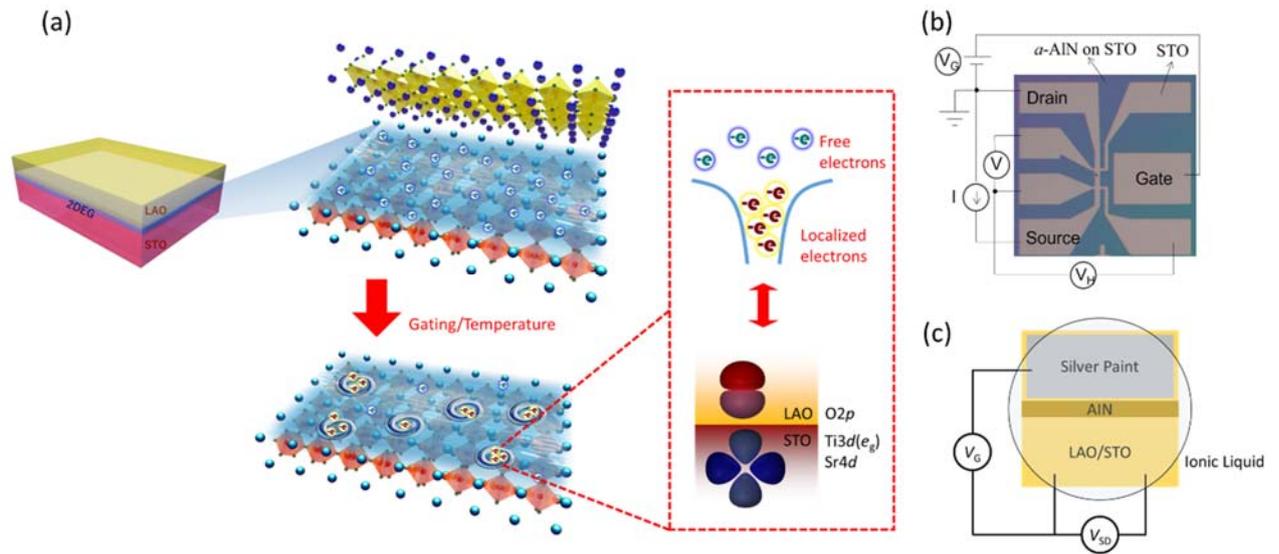

**Fig. 1.** (a) Schematic depicting the 2D charge dynamics at the LAO/STO interface that is regulated by temperature (reversible) or IL-gating processes (irreversible). By lowering the temperature or IL-gating, the 2D charges are localized at the interfacial hybridized states with contributions from the respective constituent layers (O2$p$ orbitals from the LAO layer and Ti3$d$($e_g$) and Sr4$d$ orbitals from the STO layer). (b) Optical micrograph of the device measurement circuit. (c) Schematic of large-area samples without pattern for XAS measurements. XAS measurements performed on the half without silver paint represent results after IL-gating, and that covered by silver paint for results before IL-gating. The sample was separate into two parts before any XAS measurements.

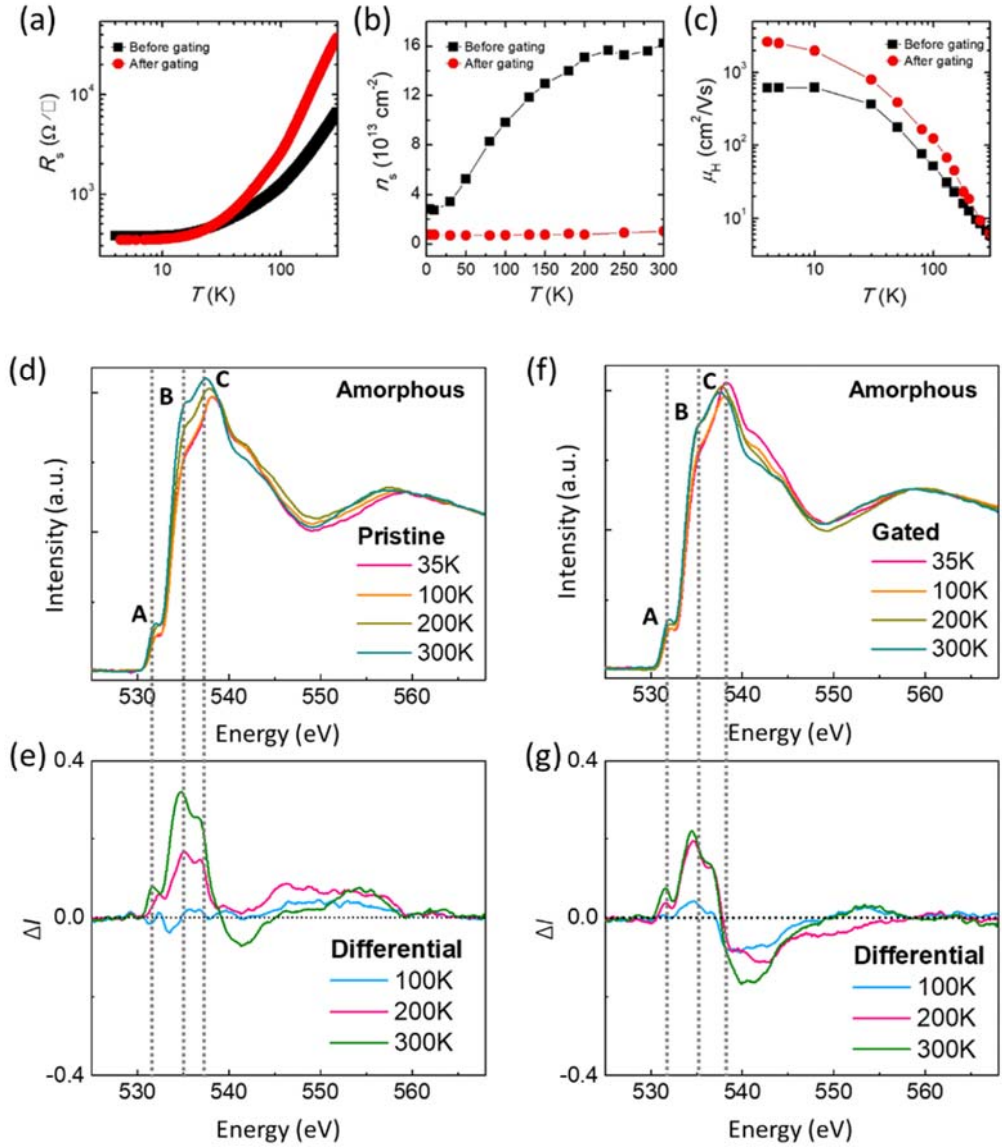

**Fig. 2.** (a) The sheet resistance, $R_s$, (b) carrier density, $n_s$, and (c) carrier mobility, $\mu_H$, as functions of temperature, $T$, for the 4.0nm-LAO/STO sample before (pristine) and after the IL-gating process has been performed. (d) Temperature-dependent O K-edge of amorphous 4.0nm-LAO/STO in its pristine state, and (e) its temperature-dependent intensity differential, $\Delta I$. Where $\Delta I = I(E,T) - I(E,35K)$. (f) Temperature-dependent O K-edge of amorphous 4.0nm-LAO/STO after ionic gating has been performed on it, and (g) its temperature-dependent intensity differential, $\Delta I$.

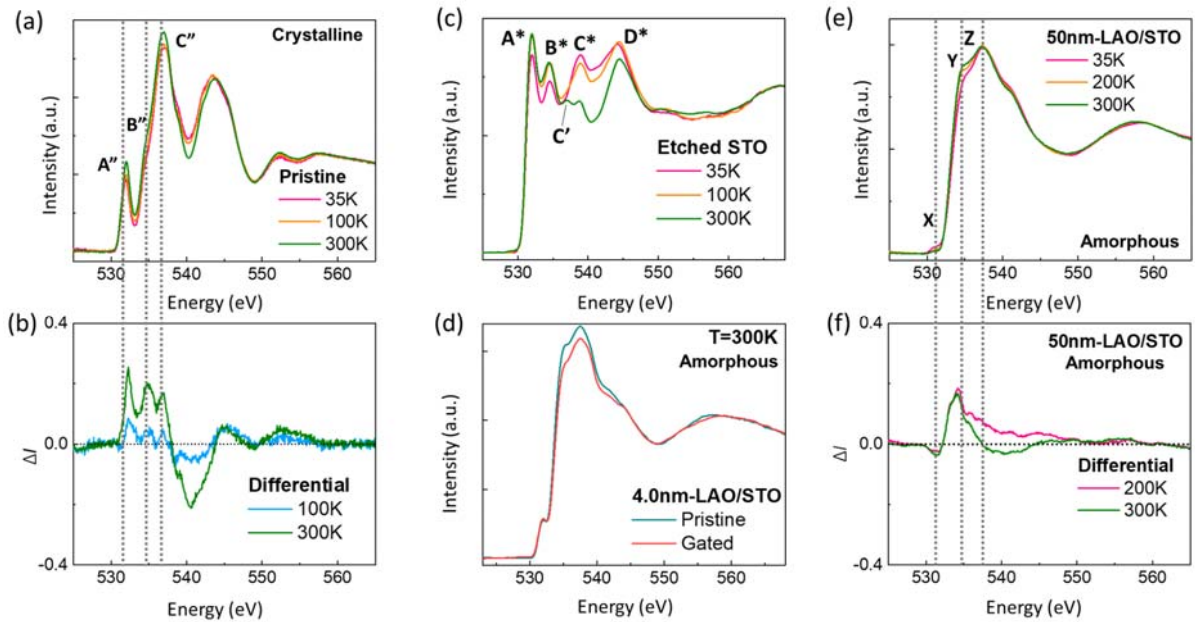

**Fig. 3.** (a) Temperature-dependent O K-edge of crystalline 4.0nm-LAO/STO in its pristine state, and (b) its temperature-dependent intensity differential, $\Delta I$. Where $\Delta I = I(E,T) - I(E,35K)$. (c) Temperature dependent O K-edge of etched STO substrate. (d) Comparing the O K-edge spectra of pristine and gated 4.0nm-LAO/STO at 300K. (e) Temperature-dependent O K-edge of 50nm-LAO/STO, and (f) its temperature-dependent intensity differential, $\Delta I$.